\documentclass[conference]{IEEEtran}
\usepackage{amsmath}
\usepackage{amsthm}
\usepackage{array}
\usepackage{graphicx}
\usepackage{relsize}
\usepackage{url} 
\usepackage{footmisc}

\usepackage{caption}
\usepackage{subcaption}  
\usepackage{multirow}
\usepackage{tabularx}
\usepackage{listings}
\usepackage{tikz}
\usepackage{color}

\ifCLASSINFOpdf
\else
\fi
\ifCLASSOPTIONcompsoc
  \usepackage[caption=false,font=normalsize,labelfont=sf,textfont=sf]{subfig}
\else
  \usepackage[caption=false,font=footnotesize]{subfig}
\fi
\begin{document}
%
\title{Collective Intelligence for Smarter API Recommendations in Python}

\author{\IEEEauthorblockN{Andrea Renika D'Souza,
Di Yang,
Cristina V. Lopes}
\IEEEauthorblockN{Department of Informatics, \\
University of California, Irvine}
\IEEEauthorblockA{\{ardsouza, diy, lopes\}@uci.edu}
}


%


\maketitle

\begin{abstract}
Software developers use Application Programming Interfaces (APIs) of libraries and frameworks extensively while writing programs. In this context, the recommendations provided in code completion pop-ups help developers choose the desired methods. The candidate lists recommended by these tools, however, tend to be large, ordered alphabetically and sometimes even incomplete.  A fair amount of work has been done recently to improve the relevance of these code completion results, especially for statically typed languages like Java. However, these proposed techniques rely on the static type of the object and are therefore inapplicable for a dynamically typed language like Python.

In this paper, we present PyReco, an intelligent code completion system for Python which uses the mined API usages from open source repositories to order the results based on relevance rather than the conventional alphabetic order. To recommend suggestions that are relevant for a working context, a nearest neighbor classifier is used to identify the best matching usage among all the extracted usage patterns.

To evaluate the effectiveness of our system, the code completion queries are automatically extracted from projects and tested quantitatively using a ten-fold cross validation technique. The evaluation shows that our approach outperforms the alphabetically ordered API recommendation systems in recommending APIs for standard, as well as, third-party libraries.

\end{abstract}


%
\IEEEpeerreviewmaketitle

\section{Introduction}
\label{sec:intro}
Programming language ecosystems include several Application Programming Interfaces (APIs), some of which are part of standard libraries, while others come from third-party developers. These APIs help software developers extend the functionality of their programs and improve software quality with very little additional code. However, learning how to use these APIs can sometimes take a lot of time.  In a study with Microsoft developers, Robillard notes that most participants attribute this difficulty of learning APIs with the scarce amount of learning resources available~\cite{robillard2009makes}. 

Code completion is one of the most widely used features in Integrated Development Environments (IDEs)~\cite{murphy2006java}. Stylos and Clarke observed that developers use this feature with the main objective of writing code faster, more correctly and to explore APIs~\cite{stylos2007usability}. 

The first generation code auto-complete tools in IDEs use the static type of an object to list out all possible attributes or methods that can be invoked or triggered when a user types a ``." character. However, M{\u{a}}r{\u{a}}șoiu et al.~\cite{muaruașoiuempirical} observed that a large fraction of the recommendations produced by these early recommenders were not being accepted by the users and were being used more frequently for debugging purposes. Hence, recent research focuses towards building ``intelligent" code completion tools which order the recommendations based on relevance rather than the conventional alphabetical order. These tools aim to improve the developer's productivity and software quality by capturing the intent of a user for generating the completion lists. 

With its rapid development features, and simple and readable syntax, and powerful libraries, Python's popularity has been growing. Being a dynamic language, the traditional intelligent techniques for code completion cannot be used for recommending APIs in Python. Variables in Python are not given a type; instead, they take the type of whichever object is assigned to them during run-time. 

Current code completion tools in Python have been ineffective for API recommendations due to the following reasons:
\begin{itemize}
\begin{figure*}	
	\centering
	\begin{subfigure}[t]{0.35\textwidth}
		\centering   
		\includegraphics[width=5cm,keepaspectratio]{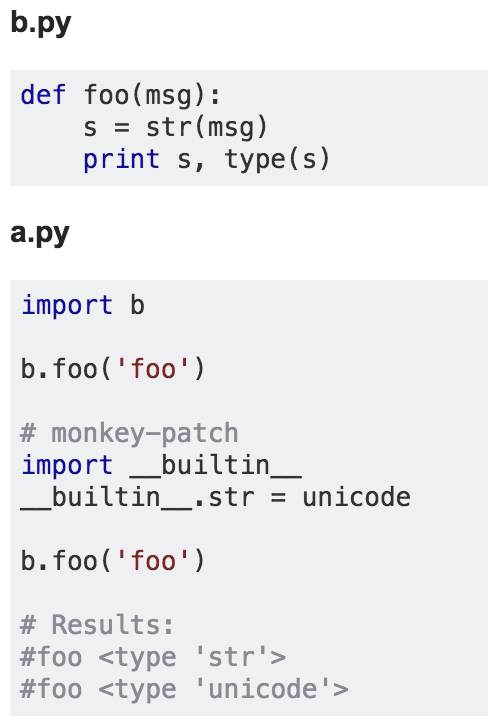}
		\caption{Monkey Patching\cite{monkeypatchexample}}
		\label{fig:monkeypatch}		
	\end{subfigure}%
	\begin{subfigure}[t]{0.55\textwidth}
		\centering
		\includegraphics[height=10cm, width=11cm,keepaspectratio]{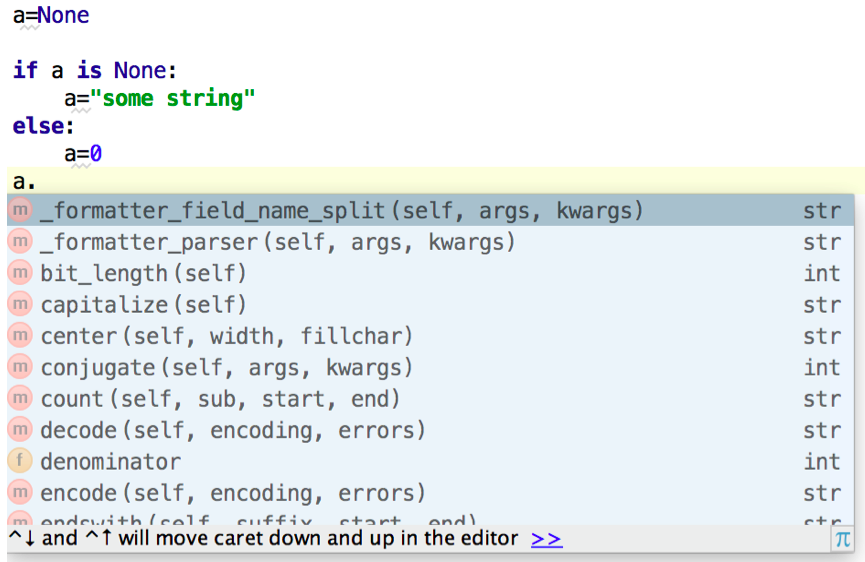}
       		\caption{Union Types in PyCharm}
   		 \label{fig:pycharm}
	\end{subfigure}
	\caption{ Examples of Dynamic behavior in Python}
    \label{fig:dynamic}
\end{figure*}

\item\textbf{Long alphabetic lists of recommendations}\\
Most code completion tools in Python display all the possible methods and attributes which can be invoked by an object, and these lists tend to be long. Some of the recommendations are rarely or never used by developers. For example, auto-complete invoked for a String object in Python retrieves around 85 results by JEDI~\cite{jedi}.  This large alphabetically ordered list makes it hard to navigate to the correct candidate, sometimes making it even slower than typing the full name of a method directly. Developers tend to rely more on prefix filtering than scrolling to reduce the number of choices~\cite{muaruașoiuempirical}. 

\item\textbf{Scarce or ambiguous documentation of APIs}\\
API documentation can be ambiguous, especially in explaining the type of the object returned by these APIs. This problem is complicated even further in Python since the language does not have static types. For instance, the documentation for \em urllib.open \em states that this method would return a \em file-like object\em. Developers may find this description confusing since it does not give a clear indication of what methods or attributes can be invoked on the object returned.

\item\textbf{Incomplete static analysis of libraries}\\
Static analysis tools generate stubs for libraries which are used to assist the development tools in recommending and approximating the return types of API methods. PyCharm~\cite{pycharm} uses python skeletons~\cite{pyskeletons} whereas Mypy~\cite{mypy} is an optional static analysis tool that uses typeshed~\cite{typeshed} for the stubs. 
These skeletons contain definitions for some of the most commonly used libraries especially the third-party ones. The list of libraries, however, is not complete since the generation of these stubs is highly dependent on the ability to perform static analysis on the source code of these libraries. According to Madsen~\cite{madsen2015static}, it is complicated to statically analyze libraries due to the following reasons:
(a) Software libraries may be partially or fully implemented in another programming language;
(b) The source code for libraries may be large and not  available for static analysis; and
(c) Dynamic features may be used in the source code of these libraries.
An absence of recommendations can, however, cause developers to suspect the presence of an error in the program or to check the additional documentation available and thus negatively impact their productivity \cite{muaruașoiuempirical}.

\item[(4)]\textbf{Failure to detect dynamic behavior}\\
In a dynamic language, it is possible that a variable can have a set of values at a particular point in the program. For example, using Monkey Patching~\cite{monkeypatch}, APIs of modules can be extended or modified using dynamic binding. In Figure~\ref{fig:monkeypatch}, the built-in function \textit{str} is modified to return a string object of type \textit{unicode} instead of \textit{str}. Some code completion systems may fail to recognize these ``union types" for an object. For example, in the code snippet shown in Figure~\ref{fig:pycharm}, the type of object named \textit{a} is guessed correctly to be both \textit{int} and \textit{str} in PyCharm~\cite{pycharm}.
\end{itemize}

These issues that exist in  current code completion systems for Python sparks the need for new recommenders that have a greater understanding of developers' goals and of Python's dynamic behavior in order to suggest APIs that are more suited to the programming task.

In this paper, we present PyReco\footnote{https://github.com/Mondego/pyreco}, an API recommender that uses the extracted API call sequences from open source repositories instead of conventional type inference techniques for the purpose of code completion. The intuition behind our approach is that a large number of extracted API usage patterns present in these projects should be able to capture all the diverse scenarios in which APIs are currently being used by developers. 

The salient features of PyReco are as follows:
\begin{itemize}
\item[(1)] A maximum of ten methods or attributes are recommended for each completion query, and they are ranked based on relevance using our Nearest Neighbor classifier, Best Matching Object. 
\item[(2)] Code recommendations become possible for all libraries and APIs that were used in the mined open source projects. Thus, the completeness of this list is based on the popularity of libraries among developers rather than the ability to do static analysis on library code.
\item[(3)] Our approach for extracting the API usages leverages the semantics of the Python language and control flow information present in program to predict the dynamic behavior more accurately and capture the current working context of the developer.
\end{itemize}

The rest of the paper is organized as follows. In Section~\ref{sec:related}, we discuss the related work done to improve code completion results. In Section~\ref{sec:method}, we describe our approach to extract API usage patterns and the Best Matching Object algorithm used for ordering the auto-complete proposals. In Section~\ref{sec:eval}, the evaluation procedure and metrics are described. In Section~\ref{sec:results}, we present and discuss our experiment results and we conclude in Section~\ref{sec:concl}.

\section{Related Work}
\label{sec:related}
In the past few years, there has been fair amount of research done to improve the relevance of API recommendations by using context information, machine learning and statistical approaches.

Robbes and Lanza~\cite{robbes2008program} propose a code completion tool that uses temporal information like the program history to provide more relevant completions. On similar lines, Lee et al.~\cite{lee2013temporal} have an additional temporal dimension for evolutionary information on the code. In a collaborative work environment, they propose that such information could make development tasks easier.

The semantic or structural information in programs is most commonly used for context in recommenders. Heinemann et al.~\cite{heinemann2011recommending} claim that the identifier could be a good indicator for the methods that can be called for the  development task. For instance, an object named \textit{angle} could indicate the relevance of suggesting \textit{sine} and \textit{cosine} operations. A context sensitive completion approach by Asaduzzaman et al.~\cite{asaduzzaman2014cscc} tokenizes semantic information like keywords, method, class or interface names from the preceding lines as part of the context to improve the relevance of the code completion results. 

Hou et al.~\cite{hou2010towards} use a combination of grouping, sorting and filtering techniques to improve code completion. In grouping, the APIs are grouped on the basis of their functionality. Sorting is done based on type hierarchy and popularity, whereas filtering is done to filter out APIs that aren't public. However, these approaches require prior knowledge on the usage of each API which is unfeasible due to the drastic increase in the number of APIs in the past few years. 

Raychev et al.~\cite{raychev2014code} model the extracted method call sequences into statistical language models like N-Gram and recurrent neural networks to predict recommendations. This approach has been proven to be fast and efficient in determining the likelihood of the next method call for Android APIs. 

Bruch et al.~\cite{bruch2009learning} propose a Best Matching Neighbor (BMN) algorithm which is used to identify the nearest neighbors among the examples of API usages. These identified neighbors are then used for recommendations.  

Bayesian networks is another machine learning model that has been used to predict the next most likely method for code completion.
Proksch et al.\cite{proksch2015intelligent} use a Bayesian networks classifier along with context information to determine the likelihood of a method being invoked. These Pattern Based Bayesian Networks also incorporate clustering techniques to reduce the model size and increase efficiency. These Bayesian networks were more effective than BMN for the SWT framework in Java. McCarey et al. \cite{mccarey2006recommending} also analyze the effectiveness of using a Bayesian techniques for recommending library methods. However, the experiments show that a Vector Space Model outperforms the Naive Bayes, Bayesian Network, Tree Augmented Naive Bayes based classifiers.
 
However, all of the above proposed improvements have been implemented for a static typed language, specifically Java. The main objective of these approaches is to order the methods or attributes for an object of a particular type.

Sch{\"a}fer et al. \cite{schafer2013effective} describe an approach based on static pointer analysis for smarter code completion results in JavaScript, a dynamic language. However, this analysis is flow-insensitive and thus, may not be able to detect some of the dynamic behavior noticed in Python like union types. Also, the APIs models used in this static pointer analysis method are generated after applying dynamic analysis on the framework's test suite that is not available for most frameworks or libraries for Python. 

Franks et al. \cite{franks2015cacheca} propose CACHECA that captures the localized regularities in a program by using its recent token usage frequency. This could, however could lead to some false positives in the code suggestions for a dynamic language which is not backed by types. 

Our approach for ranking the recommendations is based on the BMN algorithm since it outperforms techniques which incorporated association-rule mining~\cite{bruch2006fruit} and statistical techniques based on usage frequency. 

\section{Approach}
\label{sec:method}
In PyReco, we extract object usages from several GitHub projects and use a nearest neighbor classifier on the extracted usage patterns to order our recommendations based on relevance.

The first phase of our implementation involves extracting the library object usages by applying static analysis on the abstract syntax trees of the python source files. Our AST Parser uses the abstract syntax trees generated by the ast \cite{ast} module in Python's standard library. For the second phase, we propose a Best Matching Object algorithm which is based on Best Matching Neighbor (BMN) algorithm~\cite{bruch2009learning}, to predict and order the next most likely methods by using the mined usage information.

\begin{figure}
\centering
\begin{lstlisting}[language=Python] 
file = `samples/sample.txt'
temp = `samples/temp.txt'
fi = open(file)
fo = open(temp, `w')
for s in fi.readlines():
    if s.strip():
        fo.write(s)
    else:
        fo.write(`\n')
        break
fo.close()
fi.close()
\end{lstlisting}
\caption{Example code snippet}
\label{fig:sourcecode}
\end{figure}

To illustrate our approach, we will use the code snippet described in Figure \ref{fig:sourcecode}. The file handling example consists of two file objects, \textit{fi} and \textit{fo}, which are created using the builtin function, \textit{open}. The example depicted shows a usage scenario wherein \textit{fo}, a temporary file is created to dump the data present in \textit{fi}. 

\subsection{Data Extraction}
\label{subsec:dataExtrac}
We used the advanced search API~\cite{search} of GitHub to extract open source projects rated with the most number of stars in Python. The number of stars in GitHub refers to the number of people watching the project. Since APIs and software libraries are prone to change or become deprecated with time, the choice of such popular projects would be advantageous as they would be more likely to be updated with API changes than a less starred one. For our experiments, we extracted around 20,000 projects from GitHub~\cite{GitHub}.

\subsection{Analysis of Python source files}
\label{subsec:parse}
In this part, we recursively walk through the nodes of the AST tree generated to analyze the python source files present in the GitHub projects. 

Some of the salient features of our analysis are as follows: 
\begin{itemize}
\item [(1)] The python source code files are parsed in a top-down fashion. The top-down parsing emulates the way forward-directed completion is done in an IDE. 
\item [(2)] The library and module information is extracted from the \textit{Import} nodes in the AST tree generated.
\item [(3)] An object assignment is added to the program's graph if it has been created using a library function.  
\item [(4)] An API method or attribute is recorded in the graph if the receiver object's assignment has been previously recorded and the object is still alive in the current scope.
\item [(5)]The object's death is marked when it is reassigned or when its scope ends.
\end{itemize}
\begin{figure}
\centering
\includegraphics[height=10cm, keepaspectratio]{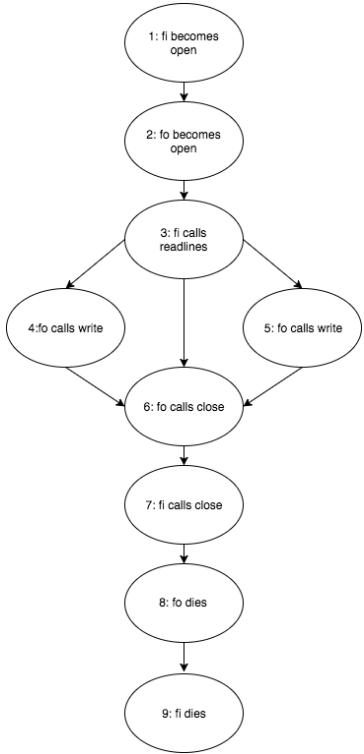}
\caption{Program Graph generated for code snippet in Figure \ref{fig:sourcecode}}
\label{fig:graph}
\end{figure}
\begin{table}[t]
\centering
\resizebox{9cm}{!}{
\begin{tabular}{|c|c|c|c|c|}
\hline
\textbf{No}&\textbf{Node} & \textbf{Entry Pts} &\textbf{Exit Pts}& \textbf{Reaching Defs} \\ \hline
1 & fi becomes open &  & 2 & {fi:open} \\ \hline
2 & fo becomes open & 1 & 3 & {fo:open, fi:open}  \\ \hline
3 & fi calls readlines & 2 & {4,5} &{fo:open, fi:open}  \\ \hline
4 & fo calls write & 3 & 6 &{fo:open, fi:open}  \\ \hline
5 & fo calls write & 3 & 6 &{fo:open, fi:open}  \\ \hline
6 & fo calls close & {3,4,5} & 7 &{fo:open, fi:open}  \\ \hline
7 & fi calls close & 6 & 8 &{fo:open, fi:open}  \\ \hline
8 & fo dies&7&9&{fo:open} \\ \hline
9 & fi dies&8& &  \\ \hline
\end{tabular}}
\caption{Reaching Definitions and extracted context information stored in Nodes}
\label{node_details}
\end{table}
\begin{figure*}[t]
\centering
\includegraphics[width=0.8\textwidth]{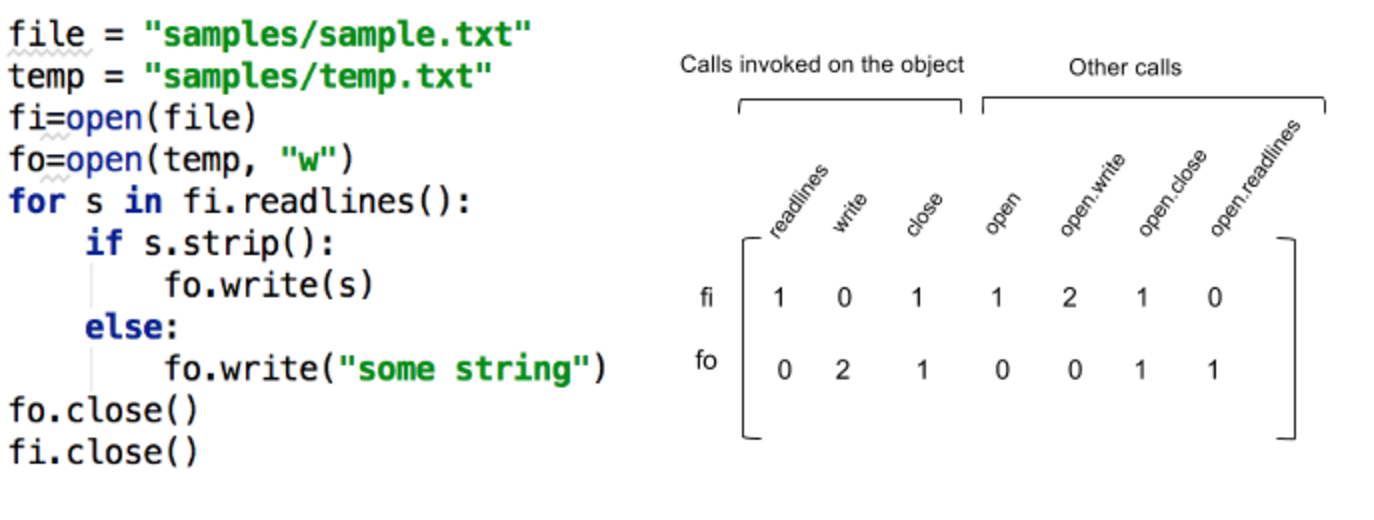}
\caption{Encoding the file objects as frequency vectors}
\label{vectors}
\end{figure*}

\begin{figure*}[t]
\centering
\includegraphics[width=0.8\textwidth]{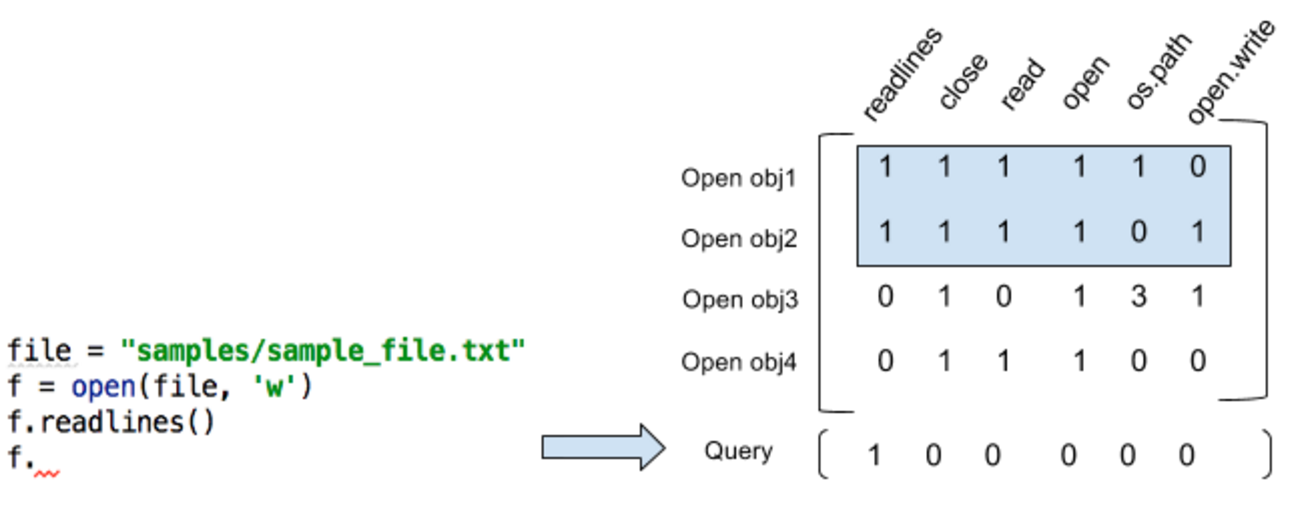}
\caption{Best Matching Objects from extracted Object Usages}
\label{bmo1}
\end{figure*} 
\begin{figure*}[t]
\centering
\includegraphics[width=0.7\textwidth]{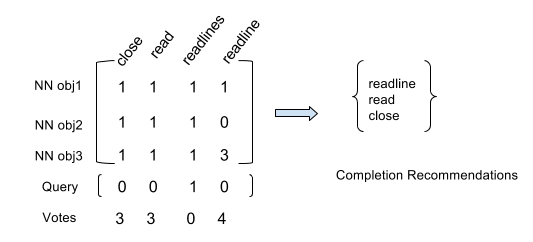}
\caption{Code Recommendations using Nearest Neighbors}
\label{bmo2}
\end{figure*} 

We use graphs in our program analysis approach since it can be used to describe the assignments and calls in terms of the control flow of a program. The program's graph splits when a branching or looping construct is encountered and merges on exiting that block. This splitting and merging of flows are depicted in the graph as shown in Figure~\ref{fig:graph}. In the example shown, the graph splits on encountering the \textit{if-else} block and the \textit{for} loop in the program. These flows join after exiting the scope of these blocks.

To describe the control flow of the program, each node contains information on its entry and exit nodes as shown in Table~\ref{node_details}. These entry and exit nodes help in the traversing across all the execution flows of the program and could be used to approximate the set of values an object can have at different points of the program using Reaching Definition Analysis~\cite{nielson2015principles}. 

Reaching Definitions analysis is done to determine all the definitions that reach a particular point in the program. At a node \textit{S} in the graph, the reaching definitions \textit{S} is the union of the reaching definitions from the entry nodes minus the ones killed at S (if an object dies at S) plus the definitions that are added in S (if an object is reassigned at S)~\cite{Reach7:online}. For instance, at node 6 (fo calls close) in Figure~\ref{fig:graph}, the reaching definition is a union of the definitions at 3, 4 and 5. No definition is added or subtracted at this node since the node at 6 marks a \textit{close} call, not the  death or assignment of an object.

These reaching definitions can be used to detect ``union types". For instance, if \textit{fo} was assigned to \textit{os.path(file)} in the else block, the reaching definitions at 6 for object \textit{fo} would be a set of values containing \textit{os.path} and \textit{open}. Thus, we could recommend methods that are invoked on \textit{os.path} as well as \textit{open}.

Currently, our approach tracks assignments made using the assignment operator (`=') or using `with' construct. An assignment node is added to the graph by evaluating the right side of the assignment expression for a library call. Certain assignments such as that of the iterator in for loops is ignored since there is no substantial information on the type of the object. For instance, the object \textit{s} in the for loop shown in Figure~\ref{fig:sourcecode} shows that it is a part of the iterator object \textit{fi.readlines()} but the statement does not clearly indicate the type of \textit{s}. 

We are able to check ``monkey patching" in most cases, by evaluating the left side of the assignment expression to check if a library method has been modified or overridden. Such calls are tracked and ignored while adding nodes to the program graph.

The program graph created after parsing the syntax trees is then used to extract call sequences for training the recommendation models. In the Figure~\ref{fig:graph}, the extracted call sequence for \textit{fo} consists of \textit{write} and \textit{close} methods, whereas for \textit{fi}, the call sequence consists of \textit{readlines} and \textit{close} methods. 

\subsection{Best Matching Object Algorithm}
After extracting the object usages from all the GitHub \cite{GitHub} projects, we train our models using a nearest neighbor classifier, which is based on the Best Matching Neighbor Algorithm~\cite{bruch2009learning}. We name this algorithm as Best Matching Object since it uses the contextual information specific to the object to recognize the nearest neighbors. 

Our tailorings in this approach are as follows:
\begin{itemize}
\item [(1)] Vectors are created for the training objects and query based on their invoking method-call frequency. 
\item [(2)] Manhattan distance is used as the distance criteria to select the nearest neighbors. 
\end{itemize}

The basic approach involves creating vectors for the object usages extracted from the projects, computing the Manhattan distance with the frequency vector created for the query, and selecting the objects with the minimum distance from the query vector as the Nearest Neighbors. The recommendations are then presented based on the decreasing order of frequencies of methods invoked by the nearest neighbors.

The calls to different methods tend to follow patterns or chain sequences in a programming task. The object's call sequence is always present in these frequency vectors generated since they indicate about an object's usage pattern. The call sequences for each object are recorded from the object's assignment nodes to the nodes marking its death through a traversal of the program graphs generated. 

As part of additional context, all the other methods that were invoked in the period between the creation and death of the object are recorded. These method calls are labeled as ``other calls'' in the frequency vectors shown in Figure \ref{vectors}.

A similar approach is followed while creating the frequency vector for the query during the evaluation process. A backward traversal of the graph is done to retrieve all the information related to the query object.
 
\begin{figure*}[t]
\centering
\includegraphics[width=\textwidth]{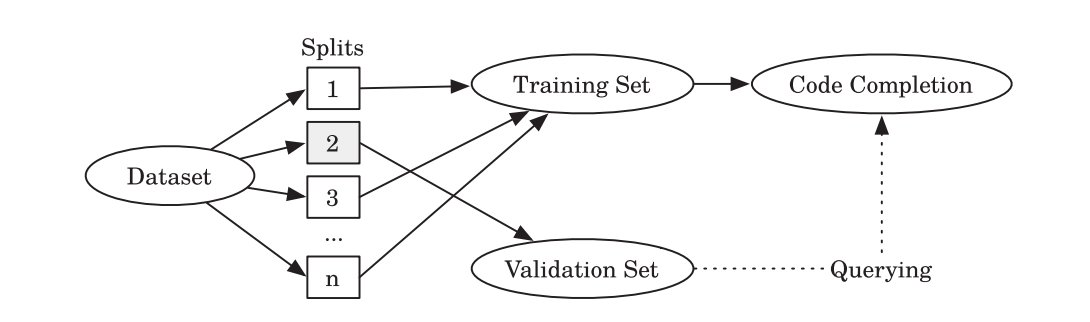}
\caption{Evaluation using Cross Validation~\cite{proksch2015intelligent}}
\label{eval}
\end{figure*}

To identify the nearest neighbors for the completion query, we compute the Manhattan distance between the query vector and other similarly defined objects found in our training dataset. The distance measure is calculated by taking the sum of the absolute values of the differences between the variables of the two vectors as shown in the following formula:

\[ \scalebox{1.5}{$d = \sum_{i=1}^{n}\left | x_{i} - y_{i} \right |$} \]

Manhattan Distance~\cite{krause2012taxicab} was selected as the distance measure for measuring the similarity between the feature vectors since it performed much better than other measures like Euclidean distance in our initial experiments.

To calculate the Manhattan distance we considered only the difference in frequency between variables present in the query vector. For instance, in the example shown in Figure ~\ref{bmo1}, only the difference in the frequency of \textit{readlines} is computed for the \textit{open} objects. Thus, n in the distance equation refers to the number of methods in the query's frequency vector with a usage frequency greater than 0.

Unlike the Best Matching Algorithm described by Bruch et al.~\cite{bruch2009learning}, we kept the original frequency values instead of reducing them to boolean factors since some methods tend to be called more frequently than others. 

In Figure 
~\ref{bmo1}, \textit{open obj1} and \textit{open obj2} are selected as the nearest neighbors since they have the minimum Manhattan distance among the extracted \textit{open} objects. The nearest neighbors then vote based on their method frequencies. This vote is done subtracting the method frequencies already present in the query vector.  The methods that are recommended in the example query shown in the Figure~\ref{bmo2} are \textit{readline}, \textit{read} and \textit{close}.

\section{Evaluation}
\label{sec:eval}
In this section, we will describe the process, metrics and the dataset that were used for measuring the effectiveness of the code recommenders.

\subsection{Experiment Procedure}
\label{subsec:procedure}
To compare the prediction quality of PyReco to other Python code recommenders, we propose an evaluation procedure consisting of automated case studies that are extracted from the GitHub projects and manual case studies from the library's official documentation and examples.

Most IDE tools depend on qualitative techniques and user studies to evaluate their effectiveness. However, it is very hard to find a representative set of users, and the approach could be time-consuming, costly and could result in subjective judgments~\cite{proksch2014towards, bruch2008evaluating}. Automating this evaluation process could give a more objective idea of the performance of code completion systems.

Our automated evaluation approach is based on the hold-out validation process proposed by Bruch et. al~\cite{bruch2008evaluating}. The automated cases were tested using a 10-fold cross-validation that consists of the following main features:
\begin{itemize}
\item [(1)] In each fold, 10\% of the library objects are kept aside as the validation set while the other 90\% are used for training.
\item [(2)] The validation set varies with each fold of the cross-validation and queries for each fold are selected randomly. The intra-project completion queries were assigned the same fold to avoid the positive bias and strong correlation among the completion queries noticed in our initial experiments and prior research \cite{proksch2014towards}. 
\item [(3)] The recommenders were queried for every method call made by these library objects by calling their code completion API, thus making the process completely automated. The program snippet till the "." character at which the method invoked is passed as input to the API. 
\item [(4)] After each query, the evaluation metrics were calculated with the relevant set containing the method the user had originally used in the program. Thus, the relevant set in our approach contains only one method.
\end{itemize}

JEDI \cite{jedi}, provides an API that can be used for retrieving the code completion results. This API is ideal for testing code completion using the above-mentioned automated approach. JEDI is a static analysis tool for Python which has been integrated into IDEs such as Atom. It has also been provided as a plugin for text editors such as Sublime. The code completion results by JEDI are alphabetically ordered.  

Manual case studies to test code recommenders usually consist of a few code completion scenarios that are identified and validated by experts~\cite{proksch2014towards}. However, in the absence of such expert validated scenarios for Python, we tested each library with a code example found in its official documentation or learning resources~\footnote{\url{https://github.com/Mondego/pyreco/tree/master/manual_queries}}. 

In these manual cases, we evaluated how PyReco fares in comparison to PyCharm \cite{pycharm}, IntelliJ's plugin for Python, which has a much more popular and powerful code completion feature than JEDI; however since their implementation is closed and it cannot be used for our automated experiments. 

\subsection{Experiment Dataset}
\label{subsec:dataset}
In the experiments proposed, the recommenders were evaluated for 20 Python libraries: 11 standard and 9 third-party. These libraries were chosen due to their popularity in Python's development community and the large numbers of library objects found in our dataset of software projects. 

\subsection{Evaluation Metrics}
\label{subsec:metrics}
To measure the quality of predictions made by the code recommenders, we used Mean Reciprocal Rank(MRR) and Recall.

Mean Reciprocal Rank is calculated by averaging the reciprocal of the rank at which the first relevant document was found across all the information needs \cite{craswell2009mean}. For a set of code completion queries \(Q\), the Mean Reciprocal Rank is defined as:

\begin{equation*}
MRR  = \frac{1}{|Q|} \sum_{i=1}^{|Q|} \frac{1}{\text{rank}_i}
\end{equation*}

Since this measure captures the rank of the relevant result, it does not penalize the systems that retrieve long lists as \(Precision\) does. MRR is also more effective in measuring the prediction quality in a ranked retrieval scenario as compared to the other set-based measures. In our case, this measure becomes equivalent to \(Mean\ Average\ Precision\ (MAP)\) since our relevant set contains only one element, the method that the developer used in the program from which the query was generated. 

We use \(Recall\) to estimate the ``completeness" of our results. It gives us an idea of the number of times the recommendation list did contain the method that the user was looking for. \(Recall\) \cite{nielson2015principles} is calculated using the following formula:

\begin{equation*}
Recall = \frac{\left | \left \{ relevant\ documents \right \} \bigcap \left \{ retrieved\ documents \right \}\right |}{\left |  \left \{ relevant\ documents \right \}\right |}
\end{equation*}

Among the measures described, we considered \(MRR\) as the main criteria to decide the quality of the results when the metrics have contrasting values.

To illustrate the way these metrics are calculated, consider the code recommendations (\textit{readline, read, close}) made in Figure~\ref{bmo2}. If \textit{read} was the method  the user used for the query shown in Figure~\ref{bmo1}, then the values of the metrics will be as follows:
\begin{enumerate}
\item [(1)] \(MRR=0.5\) since the rank at which \textit{read} was found is 2.
\item [(2)] The \(Recall\) for this query is 1.0 since \textit{read} was in the set of retrieved recommendations.
\end{enumerate}

\begin{table*}[t]
\centering
\resizebox{10cm}{!}{
\begin{tabular}{|c|c|c|c|c|}
\hline
\textbf{Library} & \textbf{PyReco-MRR } &\textbf{JEDI-MRR}& \textbf{PyReco-Recall}&\textbf{JEDI-Recall}\\ \hline
os & 0.592 & 0.037 & 0.943 & 0.356\\ \hline
re & 0.727 & 0.196 & 0.967 & 0.853 \\ \hline
ctypes  & 0.369& 0.146 & 0.565 & 0.161  \\ \hline
logging & 0.425 & 0.080 & 0.730 & 0.615\\ \hline
datetime&0.485&0.040 &0.845&0.429  \\ \hline
time & 0.516& 0.0068 & 0.951& 0.068 \\ \hline
json & 0.632 & 0.0137 & 0.950& 0.068  \\ \hline
collections  & 0.418 & 0.161 & 0.776 & 0.665  \\ \hline
struct & 0.646 & 0.237  & 0.927 & 0.843 \\ \hline
subprocess&0.560 &0.260 &0.925&0.741\\ \hline
argparse & 0.306 & 0.424 & 0.422 & 0.518  \\ \hline
\end{tabular}}
\centering
\caption{Results for Python Standard Libraries}
\label{standard}
\end{table*}

\begin{table*}[t]
\centering
\resizebox{10cm}{!}{
\begin{tabular}{|c|c|c|c|c|}
\hline
\textbf{Library} & \textbf{PyReco-MRR } &\textbf{JEDI-MRR}& \textbf{PyReco-Recall}&\textbf{JEDI-Recall}\\ \hline
django & 0.467 & 0.001  & 0.687 & 0.003 \\ \hline
numpy & 0.424 & 0.009  & 0.783 & 0.006 \\ \hline
mock  & 0.252& 0.000  & 0.472 & 0.000 \\ \hline
sqlalchemy & 0.551 & 0.092  & 0.871 & 0.419 \\ \hline
PyQt4 & 0.559 & 0.000 & 0.896 & 0.000 \\ \hline
theano & 0.674 & 0.000  & 0.930 & 0.000 \\ \hline
wx & 0.568 & 0.000 & 0.842 & 0.000 \\ \hline
google  & 0.638 & 0.001 & 0.910 & 0.002 \\ \hline
flask & 0.481 & 0.000 & 0.819 & 0.000 \\ \hline
\end{tabular}}
\centering
\caption{Results for Third-Party Libraries}
\label{third-party}
\end{table*}

\section{Results and Findings}
\label{sec:results}
In this section, we will discuss the results and findings of our experiments conducted to assess the effectiveness of PyReco as a code recommender. 

\subsection{Automated Queries}
In order to compare the performance of PyReco with JEDI, we sent the same set of completion queries for the standard and third-party libraries using the 10 fold cross validation technique described in Section~\ref{sec:eval}. Each recommender is tested with an average of \(30,000\) code completion queries sent across ten different splits for cross-validation. The evaluation metrics were averaged across the queries and are listed out in Table \ref{standard} and Table \ref{third-party}.

For standard libraries, PyReco has an average MRR value of 0.5 which means that on an average, the relevant result appears in the second position, whereas, JEDI has an average MRR value of 0.11, which indicates that the relevant result appears in the ninth position most of the times. The recall values show that the relevant result does not appear in more than half of the code completion queries for JEDI. Thus, in addition to the alphabetic order, the low recall value could also explain the JEDI's low MRR value since the MRR value is 0 in half the queries tested.

Among the standard libraries, we notice that the recall and MRR values for PyReco are low for \textit{mock} and \textit{argparse}. Since our static analysis captures the data-flow and call sequences concerning only the library objects, we find that these extracted sequences were not sufficient to capture the developer's working context for these libraries. For instance, predicting the relevant method for an object returned by the library function \textit{ArgumentParser.parse\_args} in the \textit{argparse}, would require more information on the iterator value being passed to this function. A \textit{mock} library object is usually used for unit-testing, thus it could have variable methods that are based on the function being tested.

These experiments also show the ineffectiveness of JEDI to recommend methods for third-party library objects. JEDI fails to propose auto-complete suggestions for the code completion queries in most cases. This fact is corroborated by low MRR and Recall values nearing 0. PyReco, on the other hand, has an average Recall value of 0.84 which means that the relevant result is in the ten recommendations in 84\% cases. The average MRR value for PyReco, like the standard libraries, is around 0.5. 

Among the results for the third-party libraries, we notice that the library, \textit{ctypes} has a comparatively lower recall and MRR value. In \textit{ctypes}, a CDLL object is created using a string containing the name of C library. Since our static analysis does not capture this string, the relevant result could be absent from the recommendations made by PyReco.

The low values for JEDI could be due to its inability to recognize ``union types'' and recommend for library objects that are not documented. This can be observed in the high recall values for \textit{re}, \textit{struct} and \textit{subprocess} as compared to the third-party libraries.

The results for standard and third-party libraries show that PyReco does outperform JEDI in terms of prediction quality and completeness of recommendations by a huge margin. However, the results of this experiment can not be used to explain the impact of a relevance based ordering since JEDI fails to recommend methods for most of the completion queries. The results, however, does indicate that PyReco captures the developer's context and approximates the query object's type much more effectively than JEDI does. 

\begin{table}[t]
\centering
\resizebox{5.5cm}{!}{
\begin{tabular}{|c|c|c|}
\hline
\textbf{Library} & \textbf{PyReco} &\textbf{PyCharm} \\ \hline
django & 1 & 25  \\ \hline
os & 1 & 7  \\ \hline
numpy & 1& 51  \\ \hline
re & 1& 4  \\ \hline
mock & 0 & 0\footnotemark[3]   \\ \hline
ctypes  & 2 & 0\footnotemark[3] \\ \hline
logging & 8& 3  \\ \hline
sqlalchemy&5&3 \\ \hline
datetime&1&22   \\ \hline
time & 7& 0  \\ \hline
PyQt4 & 8 & 0   \\ \hline
theano & 3 & 7  \\ \hline
wx & 1 & 57  \\ \hline
json & 2& 9   \\ \hline
collections  & 1 & 7  \\ \hline
struct & 2 & 6  \\ \hline
subprocess&1&28\\ \hline
google&2& 0\footnotemark[3] \\ \hline
flask  & 3 & 95  \\ \hline
argparse & 2 & 55 \\ \hline
\end{tabular}}
\centering
\caption{Ranks for the relevant method in PyReco and PyCharm}
\label{ranks}
\end{table}
\footnotetext[3]{no recommendations were made for the completion query}
\subsection{Manual Queries}
Since JEDI was not very effective in predicting methods for third-party libraries, we conducted experiments with PyCharm to assess the effectiveness of a relevance-based ranking as compared to the conventional alphabetic order. 

For each of the above-mentioned 20 libraries, we tested recommender with an example found in the library's online documentation or learning resources. The rank at which the relevant method was found in completion pop-up was recorded as shown in Table \ref{ranks}.

In 16 of the 20 cases tested, the ordering of the recommendations by PyReco was found to be better than PyCharm. Since recommendations made by PyReco are on the first page of the pop-up, it could positively impact a developer's productivity.

We noticed that in 7 examples, the relevant result was not found on the first page of PyCharm's auto-complete pop-up. On the other hand, PyReco places the relevant result in the first or second position in these seven examples. 

For libraries like \textit{mock}, \textit{ctypes} and \textit{google}, PyCharm fails to provide any recommendations. This inability could be due to the absence of stubs or skeletons for these libraries.

The results from these manual cases studies do indicate that a relevance based ordering like the one in PyReco could be more useful for recommending APIs.

\subsection{Threats to Validity}
We identified the following threats in our described approach:
\newline 
\begin{enumerate}
\item [(1)]\textbf{Generalization based on tested libraries}\newline
Our results are summarized based on our experiments using 20 library methods. However, these findings could be challenged when other libraries are used. Our choice of the libraries was based on the high frequency of usage patterns detected in our dataset of projects and its popularity in the Python development community.
\item [(2)]\textbf{Generalization based on manual evaluation}\newline
Our results for the manual queries are summarized based on a single query. A single query may not be representative of the performance of a library. These queries were only used to evaluate how PyReco fares against PyCharm since an automated evaluation was not feasible in the latter.

\item [(3)]\textbf{Presence of bugs in repositories}\newline
There is a possibility that some of the source code files used for training may contain bugs and thus may lead to some false positives in the recommendations.  However, since the projects extracted are the top starred ones, the presence of these bugs is expected to be negligible.
\end{enumerate}

\section{Conclusion and Future Work}
\label{sec:concl}
In this paper, we described PyReco, a code recommender, to help software developers explore APIs of libraries and frameworks more effectively and efficiently in Python. 

The proposed recommender addresses the challenge of recommending APIs in a dynamic language by reusing the intelligence found in open source repositories on API usages and suggests method calls that are ordered by relevance, unlike the other code completion tools currently available for Python.

Our experimental results show that the predictions made by PyReco are much more precise and complete than those made by JEDI \cite{jedi}. They also show promise when compared with PyCharm \cite{pycharm} for the standard and third party libraries tested. 

In our experiments, we noticed capturing the call sequences was not sufficient to capture the developer's working context for libraries like \textit{ctypes}, \textit{mock} and \textit{argparse}. To improve the auto-complete predictions for these libraries, we could extract more information on the values of the primitive objects present in the program.

We plan to integrate PyReco as a plugin in the Integrated Development Tools (IDEs) for Python. However, in order to do that, we need to conduct user studies with software developers to assess the usefulness of such a plugin and for directions in its design.

\section*{Acknowledgments}
This work was partly supported by a grant from the DARPA MUSE program.


%
\bibliographystyle{IEEEtran}
\bibliography{scam16}

\end{document}